\documentclass[%
reprint,
superscriptaddress,
 amsmath,amssymb,
 aps,
pre,
]{revtex4-1}

\usepackage{graphicx}
\usepackage{bm}
\usepackage[utf8]{inputenc}
\usepackage{amsmath}
\usepackage{float}
\usepackage{amsmath} 
\usepackage{booktabs}
\usepackage[x11names,table,xcdraw
]{xcolor}

\usepackage[usenames,dvipsnames]{pstricks}
\usepackage{epsfig}
\usepackage{pst-grad} 
\usepackage{pst-plot} 

\usepackage[normalem]{ulem}

\newcommand{\pc}{{\phi^4}}

\begin{document}
\preprint{APS/123-QED}

\title{Magnetic domain wall creep and depinning: a scalar field model approach}

\author{Nirvana B. Caballero}
\affiliation{CONICET, Centro At\'omico Bariloche, Av. Bustillo 9500, 8400 S. C. de Bariloche, R\'{\i}o Negro, Argentina.}

\author{Ezequiel E. Ferrero}
\affiliation{CONICET, Centro At\'omico Bariloche, Av. Bustillo 9500, 8400 S. C. de Bariloche, R\'{\i}o Negro, Argentina.}

\author{Alejandro B. Kolton}
\affiliation{CONICET, Centro At\'omico Bariloche, Av. Bustillo 9500, 8400 S. C. de Bariloche, R\'{\i}o Negro, Argentina.}
\affiliation{Instituto Balseiro, Univ. Nac. Cuyo - CNEA, Av. Bustillo 9500, 8400 S. C. de Bariloche, R\'{\i}o Negro, Argentina.}

\author{Javier Curiale}
\affiliation{CONICET, Centro At\'omico Bariloche, Av. Bustillo 9500, 8400 S. C. de Bariloche, R\'{\i}o Negro, Argentina.}
\affiliation{Instituto Balseiro, Univ. Nac. Cuyo - CNEA, Av. Bustillo 9500, 8400 S. C. de Bariloche, R\'{\i}o Negro, Argentina.}

\author{Vincent Jeudy}
\affiliation{Laboratoire de Physique des Solides, CNRS, Univ. Paris-Sud, Universit\'e Paris-Saclay, 91405 Orsay, France.}

\author{Sebastian Bustingorry}
\affiliation{CONICET, Centro At\'omico Bariloche, Av. Bustillo 9500, 8400 S. C. de Bariloche, R\'{\i}o Negro, Argentina.}
\email{sbusting@cab.cnea.gov.ar}


\begin{abstract}
Magnetic domain wall motion is at the heart of new magneto-electronic technologies and hence the need for a deeper understanding of domain wall dynamics in magnetic systems.
In this context, numerical simulations using simple models can capture the main ingredients responsible for the complex observed domain wall behavior. We present a scalar-field model for the magnetization dynamics of quasi-two-dimensional systems with a perpendicular easy axis of magnetization which allows a direct comparison with typical experimental protocols, used in polar magneto-optical Kerr effect microscopy experiments. We show that the thermally activated creep and depinning regimes of domain wall motion can be reached, and the effect of different quenched disorder implementations can be assessed with the model. In particular, we show that the depinning field increases with the mean grain size of a Voronoi tessellation model for the disorder.
\end{abstract}

\maketitle

\section{Introduction}

The study of field-induced magnetic domain wall motion in thin ferromagnetic films has received great attention during last decades.
Basic research allowed for the promise of new technological developments relying on the motion of domain walls~\citep{RoadMap, Hellman2017, Fert2017}, and received a large impulse in reward.
In particular, magnetic thin films with perpendicular anisotropy are good candidates for high-density magnetic memory devices.
One of the advantages in these systems is the narrow domain wall width, of a few tens of nanometers, and the relatively easy control of the domain wall position with external magnetic fields or electric currents~\cite{Parkin2008, Hayashi2008}.
Therefore, the prospective development of new technologies based on domain wall motion prompts to deepen the understanding of domain walls dynamics.

How a domain wall in a magnetic material moves is dictated by the interplay between the external drive, thermal fluctuations, ferromagnetic exchange which results in a domain wall elasticity, and the disorder present in the sample.
The external force acting over a domain wall can be generically considered to be the result of the application of an external magnetic field favoring the growth of one of the domains separated by the wall.
When the magnetic field is small, domain wall motion is strongly hindered by the disorder.
The velocity of the domain wall is ruled by activation:
\begin{equation}
\label{Eq:creep0}
V = V_d e^{-\Delta E/k_B T},
\end{equation}
where $\Delta E$ is a disorder dependent energy barrier, $k_B T$ the temperature energy scale (with $k_B$ the Boltzmann constant), and $V_d$ is a reference velocity corresponding to the vanishing of $\Delta E$.
The disorder energy scale depends on the external field as
\begin{equation}
\label{Eq:energy}
\Delta E = k_B T_d \left[ \left( \frac{H}{H_d} \right)^{-\mu}-1 \right],
\end{equation}
with $k_B T_d$ a characteristic disorder energy scale, $H_d$ the depinning field where the energy barrier goes to zero, and $\mu$ the creep exponent ($\mu = 1/4$ for magnetic thin films)~\citep{lemerle_domainwall_creep,chauve_creep_long,Jeudy2016}.
Equations~\eqref{Eq:creep0} and~\eqref{Eq:energy} imply the so-called creep law, $\ln V \sim H^{-1/4}$, which is valid for fields below the depinning field, $H<H_d$.
For fields just above the depinning field, $H \gtrsim H_d$, universal power-law behavior for the velocity-field response is due to the underlying zero temperature depinning transition and can be observed in the finite temperature domain wall dynamics~\citep{Gorchon2014, pardo2017}.
Above the depinning field, $H>H_d$, the flow regime is encountered, where the velocity grows linearly
with the field
\begin{equation}
V = m H,
\end{equation}
with $m$ the mobility.
The overall non-linear velocity--field response has been observed in a wide variety of magnetic materials
with its universal features characterizing creep and depinning regimes well accounted for by three parameters:
the depinning field $H_d$, the depinning temperature $T_d$ and the velocity scale $V_d=V(H_d)$~\cite{Jeudy2016,pardo2017,Jeudy2017_parameters}. 

The use of numerical models assists to account for the full domain wall dynamics.
Simple models as the elastic line in disordered media has been useful to unveil universal
features of domain wall motion~\citep{chauve_creep_long,kolton_dep_zeroT_long, bustingorry_thermal_rounding_epl,Ferrero2013}.
The approach of the elastic line has the great advantage of allowing to obtain
very precise exponents describing the systems dynamics in the elastic limit,
which connects with analytical results.
However, the purely elastic description leaves behind several experimentally well
known features of domain wall dynamics: topological defects, fingering, overhangs, bubbles,
plasticity, multi-valuated interfaces.
Even more, nucleation phenomena cannot be assessed with this approach, thus 
rendering impossible to recreate the vast majority of experimental protocols.

Besides, two-dimensional spin models, as Ising, XY, and Heisenberg, have been
adapted for the study of creep and depinning in domain wall
motion~\cite{nolle1993,drossel1998,roters1999,ZhouPRB2009,QinPRE2012,DongEPL2012}.
Such spatial models permit indeed to simulate bubble domains and domains with overhangs,
but their intrinsic periodic pinning made these models not truly realistic or comparable
to the experiments.
E.g., most simulations of driven domain walls with these approaches were done for random-field
instead of random-bond disorder type.

Moreover, micromagnetic simulations stand as a relevant technique to address material
specific properties.
They have been intensively used to capture domain walls static and dynamic features, 
particularly in low dimensions and small systems~\citep{Boulle2013, Voto2016, Pfeiffer2017, Woo2017}.
However, this approach being detailed and exhaustive, it is not always helpful to
distinguish and individualize
the essential ingredients ruling the domain walls dynamics.
On the computational side, the main disadvantage of this technique is the large amount of resources
or time needed for its simulation~\footnote{A standard tool to perform micromagnetic simulations is Mumax (http://mumax.github.io/).
The best performance reported for this software is $\sim3.5 \times 10^8$ cell updates per second in a GTX TITAN Xp.
For our implementation of the scalar field model, in a GTX TITAN Xp, we are able to update $\sim 15 \times 10^8$~cells/s.}.
Micromagnetic simulations are mainly used to study glassy domain wall dynamics close to the depinning transition,
and in most of the studies only the $T=0$~K case is considered.
However, recently this technique has also been used to address the creep regime of domain wall motion in Pt/Co/Pt
thin magnetic films~\citep{Geng2016}, where one needs to simulate extended domain walls, i.e. domain walls whose
extent is far much larger than its internal width.
Although the creep regime has been reached~\citep{Geng2016}, some features that are not fully compatible with
experimental observations have also been observed, as for example, two distinct creep regimes.

When possible, it is desirable for numerical models and methods to mimic experimental protocols.
Polar magneto-optical Kerr effect microscopy (PMOKE) is commonly used to measure domain
wall velocity~\citep{Emori2012, Ferre2013, Burrowes2013, Moon2013, Lin2016, Wells2017, Pellegren2017}.
In a typical experimental protocol, one or several nuclei are first created, which usually present a
bubble-like configuration.
Then, finite time magnetic field pulses are applied, impelling the original domains to grow.
The measured domain wall displacement is proportional to the pulse duration, thus giving a measure of
the domain wall velocity.
The insight that these experimental techniques can provide are naturally limited by several experimental
factors: the camera resolution, magnetic field pulse characteristics as maximum amplitude and minimum width,
control of the sample temperature and sample characteristics as the defect density and disorder of the
sample under study.
Therefore, having a model capable of reproducing the experimental conditions
is highly desirable and 
should allow one to reach more quantitative comparisons between experiments and simulations.

Here, we adapt a very well known model in statistical physics, a two-dimensional scalar field model with
a double well potential, to describe the phenomenology of domain wall motion in thin ferromagnetic films.
The model lays in a mesoscopic scale, between the elastic line and micromagnetic models, allowing to cover
large spatial and temporal scales while preserving a fairly detailed control of system parameters.
After presenting the model and key considerations to obtain domain wall velocities, we show that
simulated velocity field characteristics display the well acknowledge shape in both depinning and creep regimes,
including the $\mu=1/4$ creep exponent value.
Furthermore, we investigate the dependence of the domain wall dynamics under different quenched disorders,
stressing how the present model can be used to study geometrical properties of magnetic domains.

\section{Model}

We are interested in the study of magnetic domain wall dynamics in thin films with strong perpendicular anisotropy. In this kind of systems, the magnetic moment of the material is given by the time-dependent vector field $\vec{m}(\vec{\rho},\tau)$, where $\vec{\rho}$ and $\tau$ are the two-dimensional space and time coordinates, respectively. $\vec{m}(\vec{\rho},\tau)$ is constrained to point perpendicularly to the sample plane, that we are going to take as the $x-y$ plane.
When domains are nucleated in the sample, the magnetization inside domains will still point perpendicularly to the sample plane ($z$-direction), with the same magnitude as in the rest of the sample, but with a different orientation. In the domain wall region, typically much smaller than the domain region, the magnetization will change smoothly from one value of magnetization to the other. In a system with a strong perpendicular magnetic anisotropy, the magnetization's $x$ and $y$ components will be approximately zero in the whole sample, except for the domain wall region.
As the universal domain walls glassy dynamics is independent of the domain walls magnetic structure we will consider the evolution of the magnetization $z$-direction, neglecting the contribution of the remaining magnetization components.

The scalar field $\varphi(\vec{\rho},\tau)=m_z(\vec{\rho},\tau)$ will represent the value of the magnetization $z$-component, taking real values in the interval $[-1,1]$, at position $\vec{\rho}$ in the $x-y$ plane. 
This scalar field is a non-conserved variable: it may alter its value without a corresponding flux. The evolution of such non-conserved scalar field can then be modeled, in the limit of strong perpendicular anisotropy and strong damping~\cite{nicolao2007}, through
\begin{equation}
\frac{\partial\varphi(\vec{\rho},\tau)}{\partial \tau}=-\Gamma \frac{\delta \mathcal{H}}{\delta \varphi(\vec{\rho},\tau)}+\xi(\vec{\rho},\tau),
\label{eq:langevin} 
\end{equation}
where $\Gamma$ is a damping parameter, $\mathcal{H}$ is the free energy of the system that may contain different terms describing the interactions and disorder present in the system, and $\xi(\vec{\rho},\tau)$ represents an uncorrelated thermal bath modeled as a white noise, with $\langle \xi(\vec{\rho},\tau)\rangle=0$ and $\langle \xi(\vec{\rho},\tau) \xi(\vec{\rho}',\tau') \rangle = 2 \Gamma T \delta(\tau-\tau')\delta(\vec{\rho}-\vec{\rho}')$, with $T$ acting as an effective temperature~\cite{chaikin}.
Equation~\eqref{eq:langevin} is the simplest stochastic dynamical model in which a single non-conserved scalar field $\varphi(\vec{\rho},\tau)$ is in contact with a constant temperature heat bath. It has been already used in related problems such as the formation of magnetic patterns~\citep{jagla2004, jagla2005} or geometric pinning in magnetic films~\citep{Junquera2008, Marconi2011}.

We model the system free energy Hamiltonian $\mathcal{H}$ by following the modified $\pc$ model, as discussed by Jagla in Refs.~\cite{jagla2004,jagla2005}. In our implementation the model has three main contributions, $\mathcal{H} = \mathcal{H}_{loc}+\mathcal{H}_{rig}+\mathcal{H}_{ext}$, as described in the following. The local term, $\mathcal{H}_{loc}$, mimics the out-of-plane easy axis magnetization and thus favors the values $\varphi=\pm 1$. It is given by
\begin{equation}
\mathcal{H}_{loc}= \alpha \int \left( -\frac{\varphi(\vec{\rho},\tau)^2}{2}+\frac{\varphi(\vec{\rho},\tau)^4}{4} \right)d\vec{\rho},
\label{eq:Hlocal}
\end{equation}
with $\alpha$ proportional to the out-of-plane magnetic anisotropy constant.
A rigidity term discourages spatial variations of $\varphi$,
\begin{equation}
\mathcal{H}_{rig}= \beta \int \frac{|\nabla \varphi(\vec{\rho},\tau)|^2}{2} d\vec{\rho},
\label{eq:Hrig}
\end{equation}
with an intensity $\beta$ proportional to the exchange stiffness constant.
Finally, the external magnetic field is incorporated through the term
\begin{equation}
\mathcal{H}_{ext}= -H \int  \varphi(\vec{\rho},\tau) d\vec{\rho},
\label{eq:Hext}
\end{equation}
with a positive $H$ favoring the $\varphi=+1$ state.

We introduce two supplementary features to this simple model. First, we consider a prescription from the micromagnetic approach ensuring saturation of the local magnetization, which amounts to adding a saturation term $(1-\varphi^2)$ multiplying the external field $H$ (see Ref.~\cite{jagla2005} for a discussion). Secondly, we introduce structural quenched disorder by perturbing the value of $\alpha$ in the $\mathcal{H}_{loc}$ term. Instead of $\alpha$ we now use $(\alpha + \varepsilon \zeta(\vec{\rho}))$, with $\zeta(\vec{\rho})$ a short-range correlated random variable with uniform distribution in $[-1,1]$ and $\varepsilon$ the intensity of the disorder. This implementation of the disorder is compatible with the so-called random-bond disorder. The value of $(\alpha + \varepsilon \zeta(\vec{\rho}))$ is then a spatially fluctuating quantity giving the height of the two well potential, which controls the strength of the system anisotropy energy, and is a measure of the local field required to revert an isolated magnetic moment.

Using in Eq.~\eqref{eq:langevin} the Hamiltonian $\mathcal{H}=\mathcal{H}_{loc}+\mathcal{H}_{rig}+\mathcal{H}_{ext}$ with quenched disorder in the local term plus a saturation prescription, the evolution of the field $\varphi(\vec{\rho},\tau)$ is given by
\begin{equation}
\begin{aligned}
 \frac{\partial \varphi(\vec{\rho},\tau)}{\partial \tau}= &\Gamma \left ( 1-\varphi^2(\vec{\rho},\tau)\right )  [ (\alpha + \varepsilon \zeta(\vec{\rho})) \varphi(\vec{\rho},\tau)+H] \\
 & +\Gamma \beta \nabla^2 \varphi(\vec{\rho},\tau) + \xi(\vec{\rho},\tau). \\
\end{aligned}
\label{eq:phi4}
\end{equation}

In a sense, the model in Eq.~\eqref{eq:phi4} is a simplification of the phenomenological Landau-Lifshitz-Gilbert equation, which provides a widely acceptable micromagnetic description of the evolution of the local magnetic moment direction of the material. With some variations, it has been proven successful in modeling the magnetization of quasi-two-dimensional systems~\cite{jagla2004,jagla2005,nicolao2007,Junquera2008, Marconi2011}.

For simplicity, under a linear transformation Eq.~\eqref{eq:phi4} can be reduced to the form
\begin{equation}
\begin{aligned}
\frac{\partial \phi(\vec{r},t)}{\partial t}= & (1-\phi^{2}(\vec{r},t))h + [1 + \varepsilon\zeta(\vec{r})](\phi(\vec{r},t) -\phi^3(\vec{r},t)) \\
 & + \nabla^2 \phi(\vec{r},t) + \eta(\vec{r},t),\\
\end{aligned}
\label{eq:phi4-redunits}
\end{equation}
where we set
\begin{eqnarray}
\phi(\vec{r},t)&=&\varphi(\vec{x},\tau), \nonumber \\
\vec{r}&=&\frac{\vec{\rho}}{\sqrt{\frac{\beta}{\alpha}}}, \nonumber \\
t&=&\tau\Gamma \alpha, \label{eq:star-variables} \\
h&=&\frac{H}{\Gamma\alpha}, \nonumber \\
\eta(\vec{r},t)&=&\alpha \sqrt{\frac{\Gamma}{\beta}}\xi(\vec{\rho},\tau). \nonumber
\label{eq:units-transformation}
\end{eqnarray}
The last equality is imposed in order to ensure the proper correlation of the new effective temperature variable. From now on, all results will be expressed in reduced units, $\vec{r}$, $t$, and $h$.

In order to numerically solve Eq.~\eqref{eq:phi4-redunits} and obtain $\phi(\vec{r},t)$, we work with discretized time and space variables. We define a two-dimensional square grid with $L\times L$ cells. In each cell, $\phi$ has a uniform value updated at each step of the calculus.
For the time integration of the equation, we use the first-order numerical Euler method, with a time step of $0.1$ and given initial values. In order to implement the semi-implicit method to stabilize the numerical solution, we go through a Fourier transformation on the space variables, evaluating the exchange term at $t+\Delta t$ rather than at $t$. For more details on the numerical solution of Eq.~\eqref{eq:phi4-redunits} the reader may refer to~\cite{jagla2004}.

\section{Results}

In this Section, we first describe the adopted protocol and how the velocity of the domain wall is computed. Then we present results within the creep regime of domain wall motion and discuss temperature effects and fitted parameters. Finally, we present results depending on how the quenched disorder is implemented in the model.

\subsection{Domain wall velocity}

\begin{figure}[t!]
\centering
\includegraphics[width=1\linewidth]{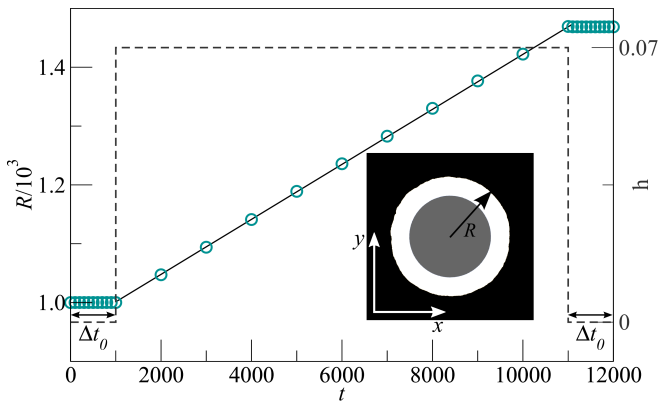}
\caption{Evolution of the effective domain radius (circles), when a field square pulse of $h=0.07$ is applied (also shown, with dashed lines) in a system at zero temperature and with a uniform disorder. The straight black line is a linear fit of the data during the application of the field pulse, which slope is indistinguishable from the obtained domain wall velocity at this field, as $\Delta R/\Delta t$ (see text). In the inset image, the spatial distribution of $\phi$ for a system with $L=4096$ cells is shown. Black color indicates the value $\phi=-1$, while gray and white correspond to $\phi=+1$. The gray circle corresponds to the initial domain (before the field pulse) and the white part is the growth of the initial domain after the field pulse.}
\label{fig:domainarea}
\end{figure}

To measure domain wall velocities we used the following protocol inspired by experiments. As the initial condition for all simulations, the scalar field $\phi(\vec{r},t)$ is set to the value $-1$ in all system cells except those cells inside a circle of radius $R_0$, centered at the middle of the system, where it takes the value $+1$.
This initial condition is then relaxed by letting the system evolve at zero field ($h=0$) for a time $\Delta t_0$ until the circle area reaches a stationary value. In order to apply an external field promoting domain wall motion, a constant field pulse of intensity $h$ is then applied during a finite time $\Delta t$. Finally, during a time $\Delta t_0'$ the system relaxes, evolving at zero field again.
In a system of size $L=4096$ with $\varepsilon=1$ in Eq.~(\ref{eq:phi4-redunits}), $\Delta t_0=\Delta t_0'=10^3$ is enough to ensure that the domain area reaches a stationary value at zero field. These parameters are kept fix at that value throughout the rest of the numerical simulations.
Note that this sequence of steps is equivalent to the sequence in which magnetic fields are applied to a sample in a PMOKE microscopy experiment, where first the sample magnetization is saturated in the $-z$-direction, a nucleation field is applied in order to generate a domain with magnetization in the $z$-direction and a square pulse is applied in order to accomplish the domain growth~\cite{Ferre2013}.

Domain wall velocities are hence computed measuring the increase in domain area during the application of the magnetic field pulse.  The area of the domain corresponding to $\phi=+1$, $A_+$, is calculated and registered during the whole simulation. Assuming a circular shape for the domains, effective radius is computed as $R=\sqrt{A_+/\pi}$. The domain velocities are then estimated as
$v=\Delta R/\Delta t$. $\Delta R=R(\Delta t_0+\Delta t +\Delta t_0')-R(\Delta t_0)$ is the effective domain radius computed as the difference between the effective domain radius before applying the field pulse and after a time $\Delta t_0'$ following the field pulse. As an example, the effective domain radius evolution for a square field pulse of intensity $h=0.07$ is shown in Fig.~\ref{fig:domainarea}.

\begin{figure}[t!]
\centering
\includegraphics[width=1\linewidth]{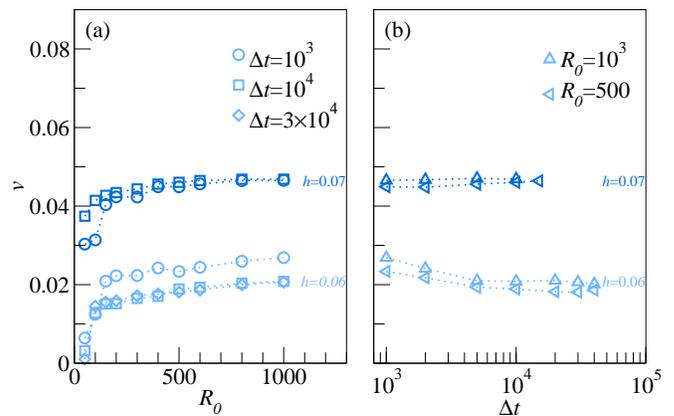}
\caption{(a) Velocity as a function of the radius of the initial domain $R_0$, for field square pulses of duration $ \Delta t$, at two values of the applied field $h$, as indicated. The initial domain radius should be large enough in order to ensure that the obtained velocities are not dependent on $R_0$. (b) Velocity as a function of the field pulse duration $\Delta t$ for two values of the applied field and two
sizes for the initial domain. Velocities may be overestimated if the pulse duration is not large enough, especially for small values of $h$.}
\label{fig:relax}
\end{figure}

In order to be consistent with PMOKE experiments, it is important that numerical results for the velocity do not depend on domain size nor pulse duration. Therefore, we check that the measured velocities are stationary and independent of the domain size. Figure~\ref{fig:relax} presents results for two values of the applied field for different initial domain sizes, $R_0$, and different durations of time pulses, $\Delta t$. We find that if $R_0$ or $\Delta t$ are too small, velocities may be underestimated or overestimated, respectively, especially for small values of $h$ close to the depinning field (see below). The underestimation of the velocities for small domain radius may be due to the domain curvature since the effective field sensed by the domain wall is corrected with a term proportional to the inverse of the domain radius ($h_{eff}=h-c/R$). This effect may not be assessed experimentally with PMOKE microscopy since it occurs at much smaller scales than the camera resolution. For instance, a typical domain wall width is $\sim$10~nm. The curvature effect according to Fig.~\ref{fig:relax} is important for $R_0\lesssim$100 simulation cells, that are equivalent to $1~\mu$m by following the transformations of Eq.~(\ref{eq:units-transformation}), with the domain wall width estimated as $\sqrt{\beta/\alpha}=$10~nm. On the other side, the overestimation of the velocities at small durations of the field pulse may be due to a memory effect of the domain walls~\cite{Ferrero2013}.
Henceforth, to ensure a representative value for the velocity, we use $R_0=10^3$ for all simulations and a carefully chosen value of $\Delta t$ for each field, in the range $10^3$ to $5\times10^6$.

\begin{figure}[t!]
\centering
\includegraphics[width=1\linewidth]{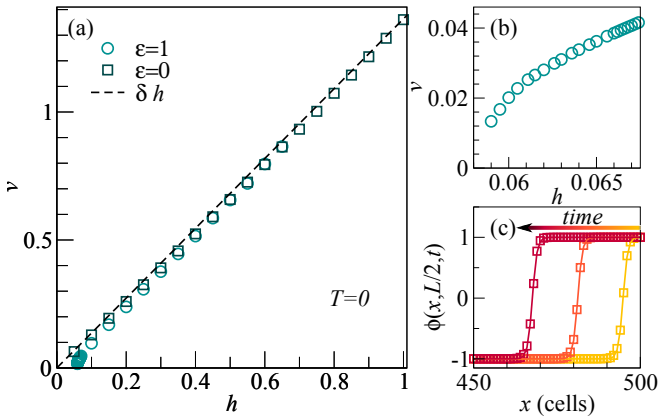}
\caption{Velocities and domain wall profiles calculated in a system at zero temperature. In (a) domain wall velocities as a function of magnetic field in a system without disorder, $\varepsilon$=0 in
Eq.~(\ref{eq:phi4-redunits}) (squares) and in a system with uniform disorder $\varepsilon$=1 (circles). The dashed black line is the linear velocity obtained from Eq.~(\ref{eq:vphi4}) with $\delta=1.4$, the domain wall width obtained from fitting domain wall profiles with Eq.~(\ref{eq:tanh}). (b) Close up view of the curve corresponding to the disordered system. In (c) domain wall profiles as a function of distance are plotted for three simulation snapshots, separated by $t=10$ in a non-disordered system for $h=1$. Fits of these curves with the function $\phi(x)=\tanh[(x-x_0)/\delta]$ are also shown.}
\label{fig:T0e0e1}
\end{figure}

When a system at zero temperature and no disorder is considered, a trivial linear behavior for domain wall velocities is found, as shown in Fig.~\ref{fig:T0e0e1} with open squares, which corresponds to a linear flow regime. The mobility $m$ of the domain wall is the proportionality factor between velocity and field and depends on its internal structure. The particular form of the domain wall, i.e. the domain wall profile, needs to be considered in order to estimate the mobility.
It is interesting to note that an estimation of domain wall velocities in the flow regime can be extracted from Eq.~\eqref{eq:phi4-redunits}. Lets consider a system of size $A$ with a $\phi=+1$ single domain of area $A_+$; correspondingly the rest of the system, $A_- = A - A_+$, has $\phi=-1$. The total system magnetization $M$ can thus be written as
\begin{equation}
 M=\frac{A_{+}-A_{-}}{A}=\frac{1}{A}\int_{A}\phi d \vec{r},
\label{eq:m}
\end{equation}
where the integral is taken over the whole system. Taking time derivatives in Eq.~\eqref{eq:m}, and using that $A = A_+ + A_-$, one obtains
\begin{equation}
 \frac{dA_{+}}{dt}=\frac{1}{2}\int_{A}\frac{\partial\phi}{\partial t} d \vec{r}.
\label{eq:areaphi}
\end{equation}
To further simplify the problem, we can consider a rectangular portion of the system, of length $l$, containing one domain wall at a position $x_0(t)$, and hence $A_+=lx_0(t)$. Under the action of an applied field $h$, the domain wall velocity can be obtained as
\begin{equation}
v= \frac{dx_0(t)}{dt}=\frac{1}{l}\frac{dA_{+}}{dt}.
\label{eq:v}
\end{equation}
Equations~\eqref{eq:areaphi} and~\eqref{eq:v} therefore relate the domain wall velocity with the time evolution of the scalar field $\phi(t)$, which is described by Eq.~\eqref{eq:phi4-redunits}. For the case of a system without disorder ($\varepsilon = 0$) at zero temperature ($T=0$), as the one considered in Fig.~\ref{fig:T0e0e1}, the velocity can be expressed in a simple form:
\begin{equation}
v= \frac{1}{2}\int \left [(1-\phi^2)(h+\phi)+\nabla^2\phi \right] dx = \delta \, h,
\label{eq:vphi4}
\end{equation}
where the integral was solved by using a functional form of the domain wall profile given by the expression
\begin{equation}
\label{eq:tanh}
\phi(x)=\tanh \left(\frac{x-x_0}{\delta}\right).
\end{equation}
For this simple model, the mobility is thus equal to the domain wall width $\delta$.
Fig.~\ref{fig:T0e0e1}(c) presents three domain wall profiles $\phi(x)$ for the direction $(x,L/2)$, taken with a time difference of $t=10$, corresponding to the case $h=1$ and without disorder at zero temperature. These profiles can be well fitted with Eq.~\eqref{eq:tanh}, giving a value $\delta = 1.4$ ~\footnote{In the $\phi^4$-model, where the double well and the elastic terms are written typically as $-\frac{1}{2}r\phi^2+u\phi^4$ and $\frac{1}{2}c(\nabla \phi)^2$ respectively, the soliton solution for the domain wall profile is $\phi(x)=\pm \phi_0\tanh[(x-x_0)/(\sqrt{2}\sqrt{c/r})]$, where $\phi_0=\pm \sqrt{r/4u}$. Since we set in our model $r=\alpha$, $u=\alpha/4$, and $c=\beta$, we obtain the solution $\phi=\pm \tanh[(x-x_0)/(\sqrt{2}\sqrt{\beta/\alpha})]$. Thus, for our model, $\delta=\sqrt{2}\sqrt{\beta/\alpha}$. In reduced units according to Eq.~\eqref{eq:units-transformation}, $\delta=\sqrt{2}\sim 1.4$.}.
In Fig.~\ref{fig:T0e0e1} we show with a dashed line the linear relationship of Eq.~\eqref{eq:vphi4} between $v$ and $h$, using $\delta = 1.4$ for the mobility, showing a fairly good agreement with the measured velocities in the so-called flow regime.

\begin{figure}[t!]
\centering
\includegraphics[width=1\linewidth]{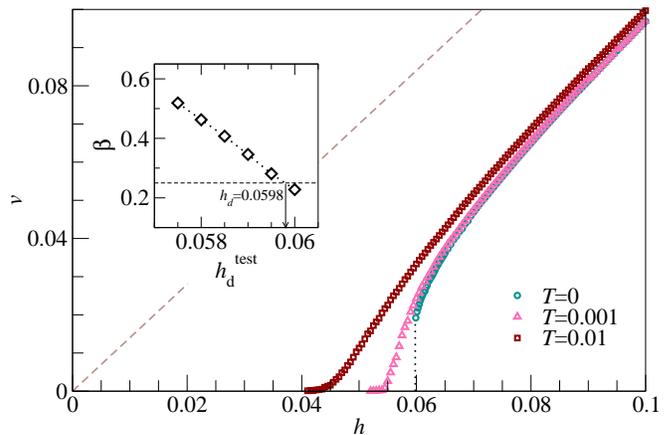}
\caption{Velocity as a function of magnetic field at three temperatures in a system with uniform disorder. Dashed line indicates the flow regime, where velocities grow linearly with slope $\delta=1.4$. The pointed vertical line indicates the depinning field $h_d=0.0598$. In the inset, $\beta$ values as a function of $h^{\mathrm{test}}_d$ are shown. The horizontal dashed line indicates the expected value $\beta=0.245$ from which the depinning field $h_d$ is estimated.}
\label{fig:vh}
\end{figure}

When disorder is considered (at zero temperature) the same linear behavior is observed at large field values, as shown in Fig.~\ref{fig:T0e0e1}(a) (circles) for a uniform disorder with $\varepsilon=1$. However, when the field is decreased the domain wall movement is strongly impeded due to the presence of disorder, resulting in a strong decrease of the velocity below $h \approx 0.06$, as shown in Fig.~\ref{fig:T0e0e1}(b). A closer inspection of this behavior is shown in Fig.~\ref{fig:vh}. At zero temperature a power-law vanishing of the velocity is expected when the depinning field is approached from above, $v \sim (h-h_d)^\beta$, with $h_d$ the depinning field and $\beta$ the depinning exponent (see Ref.~\cite{Ferrero2013} and references therein). In order to estimate the depinning field from the numerical results, one possibility is to use the method proposed in Ref.~\cite{thermal_rounding_fitexp}. With this method, from a power-law fit of the velocity against $(h-h^{\mathrm{test}}_d)/h^{\mathrm{test}}_d$, a value for the depinning exponent $\beta(h^{\mathrm{test}}_d)$ can be obtained. Based on the obtained $\beta$-values as a function of $h^{\mathrm{test}}_d$ (see the inset in Fig.~\ref{fig:vh}), the depinning field corresponds to the point where the theoretical $\beta=0.245$ value~\cite{Ferrero_PRE_2013_depinning} is reached, resulting in $h_d=0.0598$. This value is indicated with a pointed vertical line in the main panel of Fig.~\ref{fig:vh}.

\begin{figure}[t!]
\center
\centering
\includegraphics[width=1\linewidth]{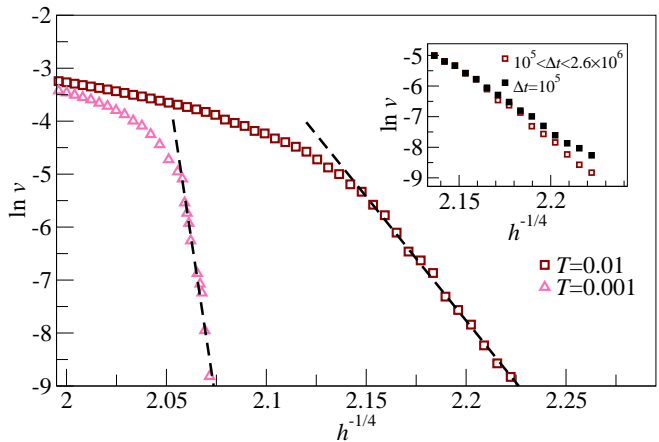}
\caption{Creep plot for a system with a uniform disorder at two different temperatures. A linear behavior, highlighted by the black dashed lines, is observed for small field values, indicating that the system is in a regime compatible with the creep regime. In the inset, data corresponding to T=0.01 is shown again with empty squares. These stationary velocities were computed from the simulation of systems where the field was applied during time lapses $\Delta t$, varying from 2.6$\times 10^6$ to $10^5$. Full squares in the inset correspond to velocities obtained at T=0.01, but with fixed $\Delta t=10^5$ and are shown in order to emphasize that some care should be taken in order to avoid the overestimation of velocities.}
\label{fig:creep}
\end{figure}

\subsection{Creep and depinning regimes}

Domain wall velocities for finite temperature values as a function of the applied field are shown in Fig.~\ref{fig:vh} for two different non-null temperatures, $T=0.001, 0.01$. Stationary velocities values are observed at fields smaller than the depinning field, $h<h_d(T=0)$, since temperature allows the activation over energy barriers, as expected in the creep regime. As indicated by Eq.~\eqref{Eq:creep0} and the field dependence of the energy barrier, a linear relationship between $\ln v$ and $h^{-1/4}$ should be observed in the creep regime. Such a creep plot is shown in Fig.~\ref{fig:creep} for the two finite temperature data sets. It shows that the numerical data is compatible with a creep exponent $\mu=1/4$ for the smaller field values. The inset of Fig.~\ref{fig:creep} shows the dependence of the velocity with the pulse duration $\Delta t$ in a creep plot, showing how the stationary velocity limit is reached at increasing $\Delta t$ for low fields. This should be carefully taken into account in numerical simulations.

\begin{figure}[t!]
\center
\centering
\includegraphics[width=1\linewidth]{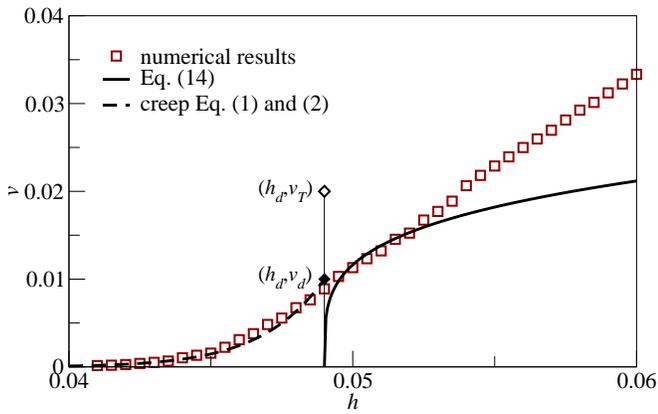}
\caption{Velocity-field curve at $T=0.01$, analyzed with the method proposed by Diaz Pardo et. al~\cite{pardo2017} for experimental curves. Dashed black line is a fit of data below the depinning field $h_d$, denoted with a vertical black line, following the creep law (Eqs.~\eqref{Eq:creep0} and~\eqref{Eq:energy}). The black continuous line indicates the curve corresponding to $v(h,T)$ just above depinning, obtained without adjustable parameters, which corresponds to the predictions of Eqs.~(\ref{Eq:depinning0}). The full black diamond indicates the point ($h_d,v_d=v(h_d$)), the upper boundary of the creep regime, while the empty diamond corresponds to $v_T=v_d (T_d/T)^\psi$ (see discussion in the text).
}
\label{fig:av}
\end{figure}

In order to discern how far one can progress on the comparison between the model and experimental results, we use the same fitting procedure as recently used for experimental data~\cite{pardo2017,Jeudy2017_parameters}. This allows one to extract the three key parameters describing the glassy dynamics within creep and depinning regimes: the depinning field $h_d$, the depinning temperature $T_d$, and the velocity scale $v_d = v(h_d)$. The fitting procedure is described in detail in Ref.~\cite{Jeudy2017_parameters}. In brief, the depinning field and the velocity scale are first estimated using the inflection point of the $v(h)$ curve, which allows one to estimate the depinning temperature from the slope of the creep plot. Then the full model, Eqs.~\eqref{Eq:creep0} and~\eqref{Eq:energy}, is fit allowing to adjust the three values. Finally a fine tunning is achieved using that, just above depinning, the velocity presents signals of the zero temperature depinning transition \footnote{Note that $v_d(T) \sim T^\psi$ when $T \ll T_d$, thus leading to $v(h,T \ll T_d^\psi) \sim (h-h_d)^\beta$ (see Ref.~\cite{pardo2017} for details).},
\begin{equation}
\label{Eq:depinning0}
v(h,T)= \frac{v_d(T)}{y_0} \left( \frac{T}{T_d} \right)^{-\psi} \left(\frac{h-h_d}{h_d} \right)^\beta,
\end{equation}
with $y_0=0.65$ a fixed universal constant and $\psi=0.15$ the thermal rounding exponent~\cite{bustingorry_thermal_rounding_epl,Gorchon2014,pardo2017}.
Results of the fit using the creep law, Eqs.~\eqref{Eq:creep0} and~\eqref{Eq:energy}, and the depinning transition scaling, Eq.~\eqref{Eq:depinning0}, to the velocity-field numerical data are plotted in Fig.~\ref{fig:av} for $T=0.01$. Summarizing, the obtained values for the depinning field are $h_d(T=0.001)=0.0558$ and $h_d(T=0.01)=0.0490$, for the depinning temperature we get $T_d(T=0.01)/T=(89\pm1)$ and $T_d(T=0.001)/T=(495\pm20)$, and for the velocity scale $v_d(T=0.01)=(0.010\pm0.005)$ and $v_d(T=0.001)=(0.0070\pm 0.0005)$. It has been shown using experimental data that values of $v_T=v_d (T_d/T)^\psi$ are expected to coincide with the velocity of the linear flow regime~\cite{pardo2017,Jeudy2017_parameters}. For our numerical model, although the fit gives reasonably good values for $h_d$ and $T_d$, the value of $v_d$ gives a value of $v_T$ far below the linear flow regime. This feature of the model is due to a large crossover between the creep and the flow regimes, that is also observed in velocity-force curves obtained with the elastic line model~\cite{bustingorry_thermal_rouding_long}. 
Overall, we have shown that the numerical data can be fit using the same fitting procedure as used to deal with experimental data.

\begin{figure}[t!]
\centering
\includegraphics[width=1\linewidth]{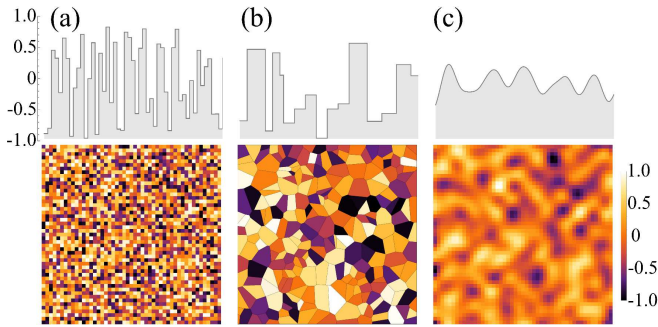}
\caption{Differences between the used disorder types illustrated with a single realization. Each pair of images show in top the value of the disorder parameter $\varepsilon$ along the first line of cells of each grid showed in the bottom: (a) uniform disorder, (b) Voronoi disorder, and (c) filtered disorder. Bottom images correspond to a portion of $50\times 50$ cells.}
\label{fig:disorders}
\end{figure}

\subsection{Models of disorder}

Finally, since the specific model of disorder is, at least partially, responsible of the domain wall dynamics, and in order to stress potential applications of the present model, we show how the velocity-curve depends on the underlying disorder model. We then study the variation of domain wall velocities using three different disorder models. In the first disorder type, already presented, the values assigned to the disorder ($\zeta(\vec{r})$ in Eq.~(\ref{eq:phi4-redunits})) were randomly chosen from a uniform distribution over the range $[-1,1]$, independently for each numerical cell in the system. For the second disorder type, we use a Voronoi tessellation of the system with $N_V=1.5\times10^6$ Voronoi grains and give for each grain a constant $\zeta$ value between -1 and 1 from a uniform distribution.
Finally, as a third disorder model, a filtered disorder, is built by using a standard low pass filter over an independent random uniform distribution. These three disorder types are shown in Fig.~\ref{fig:disorders} for a fraction of $50\times50$ cells of the two-dimensional system.

\begin{figure}[t!]
\centering
\includegraphics[width=1\linewidth]{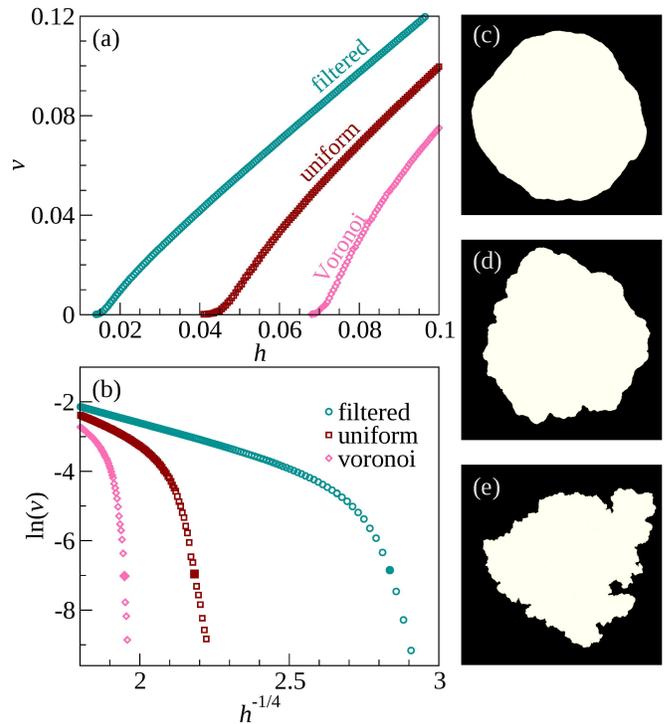}
\caption{(a) Velocity-field curves for three different disorders types at T=0.01. In (b) a creep plot is shown for the three velocity-field curves. The images, (c-d-e), show the final configuration of the domains for different fields values corresponding to similar velocities in the creep regime, as indicated by the full large symbols in (b). The shown domains correspond to systems with (c) filtered, (d) uniform, and (e) Voronoi disorders.}
\label{fig:disordersdynamics}
\end{figure}

Numerical results for the velocity-field curve for the different disorder models are shown in Fig.~\ref{fig:disordersdynamics}(a). As can be observed, velocity scales are visibly dependent on the type of implemented disorder. For a given field, velocity decreases when the considered disorder model passes from the filtered disorder to the uniform disorder and to the Voronoi disorder. In fact, the lowest depinning field is obtained for the filtered disorder, while the greatest depinning field corresponds to the Voronoi tessellation model. One can also observe that although $h_d$ changes with the disorder, and $T_d$ and $v_d$ probably too, the general shape of the velocity-field curve seems to be preserved. This means that universal features, as the critical exponents, are not presumably changing. In fact, the creep plot presented in Fig.~\ref{fig:disordersdynamics}(b) shows that the creep regime for the three different disorder models can be well described using the universal creep exponent $\mu=1/4$. The present model can also be used to investigate the effect of different disorder types on domain walls geometrical properties. Figures~\ref{fig:disordersdynamics}(c-d-e) show the shape of the domain for the three studied disorder types, all obtained at the same velocity within the creep regime, as indicated by full large symbols in Figure~\ref{fig:disordersdynamics}(b). A simple inspection shows that the roughness of the domain's shape increases with the value of the depinning field, depending on the type of disorder model used. 

\begin{figure}[t!]
\centering
\includegraphics[width=1\linewidth]{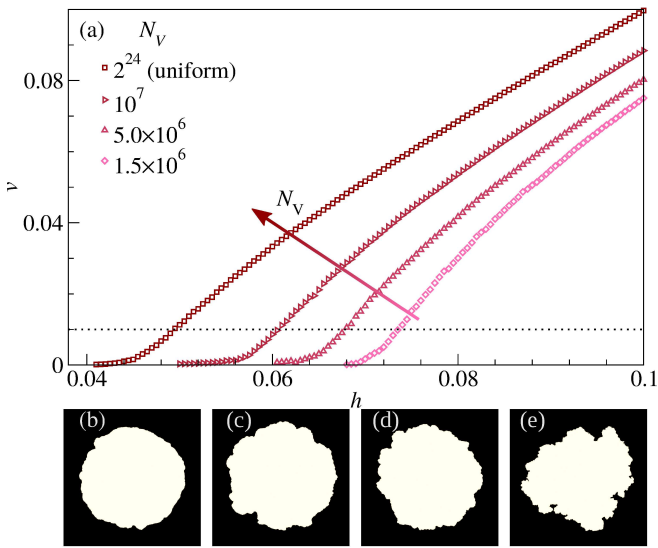}
\caption{(a) Velocity-field curves for four different numbers of Voronoi grains $N_V$ in the implementation of the disorder in a system at T=0.01. When the number of Voronoi cells is equal to the system size ($N_V=2^{24} \approx 1.7 \times 10^7$), the uniform disorder is recovered. The shown domains correspond to systems with (b) $N_V=2^{24}$, (c) $N_V=10^7$, (d) $N_V=5\times10^{6}$, and (e) $N_V=1.5\times10^{6}$, for simulations with different field values corresponding to similar velocities, indicated by the dotted horizontal line.}
\label{fig:ovhvoronoi}
\end{figure}

Until now, we showed qualitatively how different domain geometries and depinning fields may be obtained by changing the disorder implementation, but the comparison was not \textit{fair} in the sense that the uniform disorder and the filtered disorder have different correlation lengths and intensities.  On the other side, a uniform disorder can be recast as an extreme case of a Voronoi tessellation, where the smallest possible area $\sigma$ for the Voronoi grain sizes is considered. Thus, the amplitude of the disorder is not changed but the correlation length is. To explore deeper on this point, we tested two other Voronoi mean grain sizes, by generating Voronoi tessellations of $N_V=5\times10^6$ and $N_V=10^7$ cells. The four Voronoi tessellations correspond to a mean area of the grains of $\sigma=1$, $\sigma\sim$ 1.6, 3.4, 11.2, when decreasing the number $N_V$, respectively. Velocity-field curves are shown in Fig.~\ref{fig:ovhvoronoi} for the four elections of $N_V$. The main feature to highlight is that smaller depinning fields are obtained for smaller grain sizes of the Voronoi tessellation. This dependence of the depinning field with the Voronoi mean grain sizes was observed before in micromagnetic simulations~\cite{SoucailleThesis}, although we show here that the geometrical properties of domain walls also depend on the Voronoi mean grain size (Fig.~\ref{fig:ovhvoronoi}(b-e)).

\section{Conclusions}

In summary, we have presented a study of domain wall dynamics in thin magnetic films using a
versatile effective two-dimensional model.
The model can be recognized as a generalization of the $\pc$-model of statistical mechanics,
commonly used to study phase transitions and critical phenomena~\cite{chaikin}.
It includes exchange interactions, external field, effective temperature and disorder and
can be easily extended to consider dipolar interactions. 

With the aim of conciliating numerical simulations with experiments, we treated the numerical system
using the same protocol as in experiments. 
For example, the same sequence of applied magnetic field pulses was considered,
and we discussed how to obtain stationary velocity values, independent of the initial domain
size and the pulse duration.
We showed that the lowest field velocity results are compatible with the thermally activated
creep regime.
This is an important numerical milestone, it opens the possibility to study creep dynamics at large
length and time scales with a simple but realistic and material-parameters-tunable numerical model.
We showed that our numerical results are well described by critical exponents commonly used in thin magnetic systems: $\mu=1/4$, $\beta=0.245$, and $\psi=0.15$.
This suggests that, in the range of parameters explored here, the full spatiotemporal description of the domain wall is compatible with the quenched Edwards-Wilkinson universality class. A more quantitative comparison, especially viable for the depinning regime, is left as a second step, however.  In particular, a question mark is opened to know to which extent elastic depinning scaling will hold in situations where plasticity, bubbles, and overhangs become more dominant.

Furthermore, with the same fitting procedure used to analyze experimental results, we obtained values
for the key non-universal parameters needed to describe domain wall dynamics in the creep and
depinning regimes.

Different disorders were finally considered, stressing the versatility of the model. Properly modeling the disorder landscape in thin ferromagnets is key to the understanding of domain walls dynamics and its influence on materials design. The Voronoi tessellation disorder model appears as a tractable model in this direction~\cite{SoucailleThesis}. In particular, we found that within the Voronoi tessellation disorder model the depinning field and the domain wall roughness both increase with the mean size of the Voronoi grains. We expect that fitting experimental results with the presented model would provide experimental values for the parameters characterizing the disorder, such as the mean grain size and the energy scale of the disorder landscape.
Furthermore, an overall systematic exploration of disorder-type effects on phase field and micromagnetic models for domain wall dynamics is somehow missing in the field, and the approach here presented appears as a good starting point on this direction.

Prominent features of the studied model are its adaptability to realistic model parameters
and versatility to study many different experimentally-inspired protocols that may be difficult
to actually perform in the lab. 
For example, besides the domain wall dynamics, a careful study of domains nucleation for different
disorder types with varying intensity can be performed with the same model.

\section{Acknowledgments}
N.B.C. acknowledges financial support from the IDMAG team of the LPS.
N.B.C. and S.B. acknowledge discussions with Vivien Lecomte and Federico Rom\'a.
Financial support by the France-Argentina Project ECOS-Sud No. A12E03 is acknowledged.
This work was partly supported by the Argentinian projects PIP11220120100250CO (CONICET), PICT2016-0069 (MinCyt) and UNCuyo Grants No. 06/C490 and 06/C017.

\bibliography{tfinita5,paredom,phi4,dmi,paredom2}

\end{document}